\documentclass[manuscript]{aastex}

\def\4u{4U1608-522}

\slugcomment{Submitted to the astrophysical journal}

\shorttitle{Soft lags in kHz QPOs}
\shortauthors{Barret}

\begin{document}


\title{Soft lags in neutron star kHz Quasi Periodic Oscillations: evidence for reverberation?}


\author{Didier Barret\altaffilmark{1,2}}
\affil{Universit\'e de Toulouse; UPS-OMP; IRAP; Toulouse, France}
\affil{CNRS; Institut de Recherche en Astrophysique et Plan\'etologie; 9 Av. colonel Roche, BP 44346, F-31028 Toulouse cedex 4, France}
\email{didier.barret@irap.omp.eu}



\begin{abstract}
High frequency soft reverberation lags have now been detected from stellar mass and super massive black holes. Their interpretation involves reflection of a hard source of photons onto an accretion disk, producing a delayed reflected emission, with a time lag consistent with the light travel time between the irradiating source and the disk. Independently of the location of the clock, the kHz Quasi-Periodic Oscillation (QPO) emission is thought to arise from the neutron star boundary layer. Here, we search for the signature of reverberation of the kHz QPO emission, by measuring the soft lags and the lag energy spectrum of the lower kHz QPOs from \4u. Soft lags, ranging from $\sim 15$ to $\sim 40 \rm \mu s$, between the 3-8 keV and 8-30 keV modulated emissions are detected between 565 and 890 Hz. The soft lags are not constant with frequency and show a smooth decrease between 680 Hz and 890 Hz. The broad band X-ray spectrum is modeled as the sum of a disk and a thermal comptonized component, plus a broad iron line, expected from reflection. The spectral parameters follow a smooth relationship with the QPO frequency, in particular the fitted inner disk radius decreases steadily with frequency. Both the bump around the iron line in the lag energy spectrum, and the consistency between the lag changes and the inferred changes of the inner disk radius, from either spectral fitting or the QPO frequency, suggest that the soft lags may indeed involve reverberation of the hard pulsating QPO source on the disk.
\end{abstract}





\section{Introduction}
Kilohertz quasi-periodic brightness oscillations (kHz QPOs at frequency above $\sim 300$ Hz) have been observed with the Rossi X-ray Timing Explorer \citep[RXTE, ][]{bradt93} from about thirty neutron star low-mass X-ray binary systems \citep[][for a review]{klis00}. Those signals have triggered much excitements, because their frequencies match the orbital frequencies of matter orbiting very close to  the neutron star, hence may be related to strong field general relativity effects. Most of the analysis performed so far on kHz QPOs has focussed on their frequency, RMS amplitude, quality factor, relation to continuum spectral parameters and variability at lower frequency, but very little is known about their energy dependent time lags. Soft lags were however reported using early RXTE observations of the lower kHz QPOs of 4U1608-522 and 4U1636-536 \citep{vaughan98,kaaret99,markwardt00,bronzato11}. In 4U1636-536, at $\sim 830$ Hz, the soft emission in the 3.8--6.4 keV band lags the hard emission above 9.3 keV by about 25 $\mu s$ \citep{kaaret99}, a value that is consistent with the one measured from 4U1608-522 at about the same frequency \citep{vaughan98} ($\sim 27 \mu s$ at 830 Hz between the 4--6 and 11--17 keV bands). Recently \cite{avellar13} have extended on those previous analysis by a more thorough study of both sources.  

The detection of soft lags from kiloHz QPOs of neutron stars has to be placed in a broader context, where soft lags are now being routinely detected in the high frequency variability of stellar mass black holes \citep{uttley11}, active galactic nuclei \citep{demarco11,emmanoulopoulos11,cackett13,fabian13} and even recently from cataclysmic variables \citep{scaringi13}. Those lags have great significance as they probe the dynamics of the very inner regions of the accretion flows onto compact objects. They are a natural prediction of the reflection model, in which a hard X-ray source irradiates the accretion disk, producing a delayed emission with a reflection spectrum (with prominent Iron K$\alpha$ line emission), that is skewed by relativistic effects close to the central object \citep[see][for reviews]{miller07,reynolds13}. The magnitude of the lags is consistent with the light travel time between the irradiating source and the reflector. The detection of a lag in the K$\alpha$ line emission provides clinching evidence for such a scenario \citep{zoghbi12,kara13a,kara13b,zoghbi13}. 

Independently of the mechanism setting up the frequency, it is generally agreed that the QPO modulated emission arises from the boundary layer between the neutron star surface and the accretion disk \citep[][]{gilfanov03,abramowicz07}. The strongest evidence comes from Fourier frequency resolved spectroscopy showing that the QPO energy spectrum resembles the energy spectrum of the boundary layer. The QPO spectrum is significantly harder than the average continuum emission, and certainly the disk emission, also detected in the energy spectrum \citep{gilfanov03}. 

In this paper, we investigate wether the soft lags detected in kHz QPOs can be interpreted in the framework of the reverberation lags recently detected in AGNs and stellar mass black holes. For this purpose, we measure the lags and the spectral parameters of the continuum emission as a function of the lower kHz QPO frequency. We limit the present analysis to the highest quality RXTE data of 4U1608-522, showing strong kHz QPOs between 565 and 890 Hz, as to maximize the sensitivity to lag detections \citep[basically the same data used in][]{barret05}. We note that our work extends on \cite{avellar13}, who did not use data above 800 Hz for \4u\ and did not combine the lag measurements with the spectral analysis of the continuum emission. In the next section, we describe the technique used and the main results. 

\section{Data analysis}
To measure the time lags, we have first extracted events in two adjacent energy bands: 3--8 keV and 8--30 keV, using high time resolution science event data. The cross spectrum of the soft and hard band time series gives the phase difference between the two energy bands \citep{nowak99}. This phase difference is converted back into a frequency-dependent time lag: $\tau (\nu) = \Delta \phi /2\pi \nu$. In this paper, a positive lag means that the soft band lags behind the hard band\footnote{This is the opposite convention to the one that is commonly used in AGN studies, where soft lags are usually negative.}. We have computed power density and cross spectra over segments of a fixed duration $T$ and have averaged $M$ of those to produce an individual Cross Spectrum (CP$_i$). The lower kiloHz QPO frequency  ($\nu_{l,i}$) can vary by tens of Hz over the typical duration of the observation ($\sim 3000$ seconds). The analysis presented below applies only to the lower kHz QPO of \4u, and is not suited for the upper kHz QPO, which is much harder to detect (mostly because of its larger width). As to maximize the lag detection sensitivity, we wish to bin each individual CP$_i$ over adjacent frequencies, sampling the QPO profile. This implies knowing the QPO centroid frequency and its width ($w_{l,i}$, FWHM) in each CP$_i$. In order to do this, we have first reconstructed the time evolution of the QPO frequency, extracting Fourier PDS using events in a 3--30 keV reference band. The reference energy band chosen maximizes the signal-to-noise ratio of the QPO detection, and therefore enables a better determination of the QPO frequency than in the soft and hard bands. $\nu_{l,i}$ was obtained with a maximum likelihood technique \citep{barret12}. We have also shifted-and-added all the PDS to the mean QPO frequency to obtain the mean QPO width ($\bar{w}$) over contiguous segments of data. Here we have assumed $w_{l,i}=\bar{w}$ and averaged the CP$_i$ over a frequency bandwidth of twice  $\bar{w}$. As a final product of the analysis, for each observation, we have frequency dependent time lags integrated over segments of $M\times T$ second duration and frequency bandwidth $2~\bar{w}$. Hereafter, we have assumed $T=4$ sec, which provides adequate frequency resolution to sample to the QPO profile, whose width varies from $\sim 3.5$ to $\sim 10$ Hz in the data considered here. 

\begin{figure}
\plotone{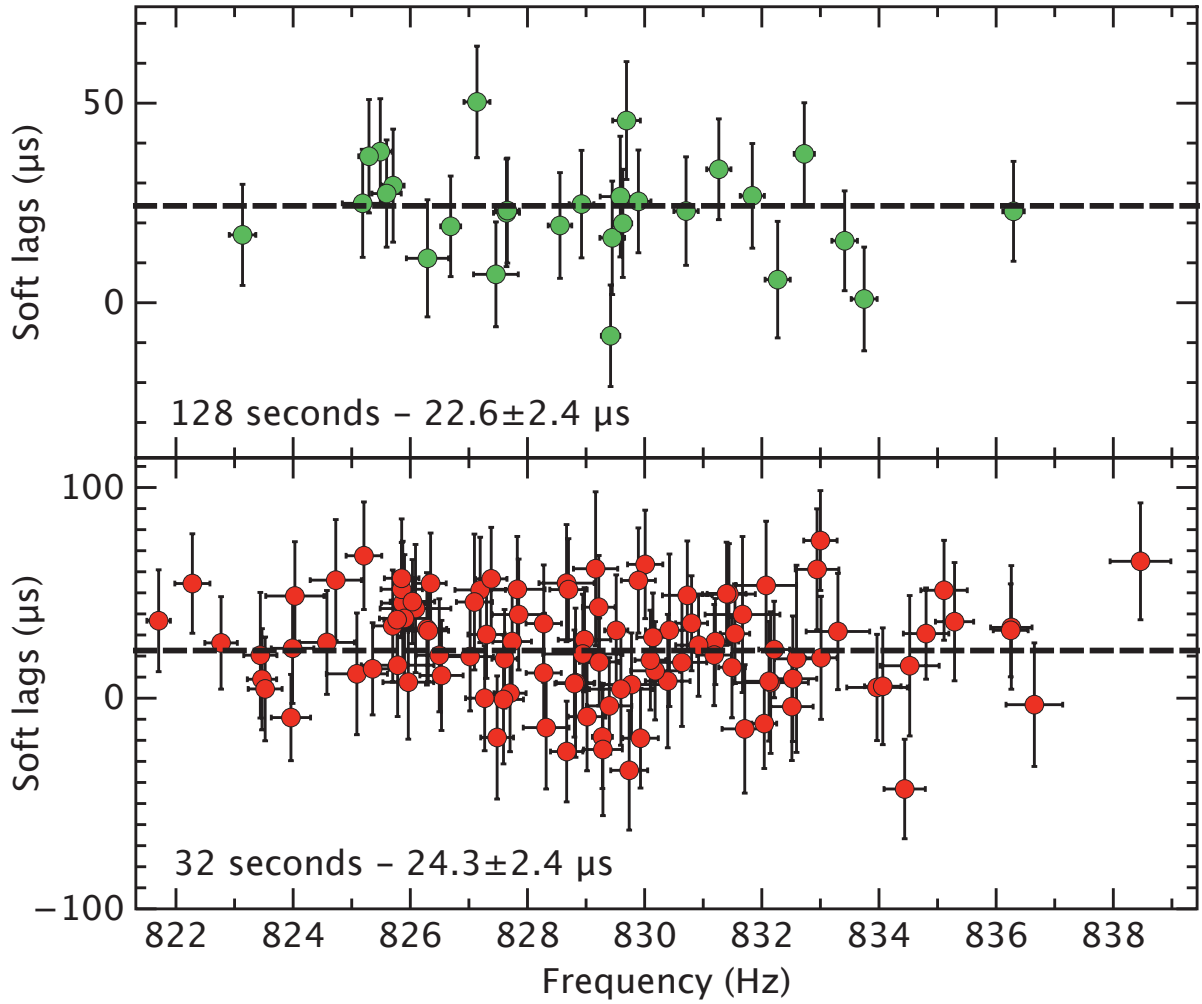}
\caption{Soft lags (in $\mu s$)  measured between the 3--8 keV and 8--30 keV bands for the second March 3rd, 1996 observation of \4u, for integration time of 32 (bottom) and 128 seconds (top panel) respectively. In both cases, the soft lags are consistent with being constant with a mean value around $\sim 23 \mu s$. The best fit value is represented by the dashed line. \label{dbf1}}
\end{figure}

In the second segment of the March 3rd, 1996 observation (ObsID 10072-05-01-00), the lower kHz QPO can be significantly detected in the soft and hard X-ray bands on  timescales as short as 32 seconds, as a combination of a large source count rate (the 5 detector units of the proportional counter array were operating) and high coherence. In Figure \ref{dbf1}, we show the soft lags between the 3--8 keV and 8--30 keV bands, measured on 32 and 128 seconds respectively (i.e. $M=8$ and 32). As can be seen, increasing $M$ yields fully consistent results, while the scatter on the lags and associated error bars have been reduced. Note that the lag is consistent with being constant in both cases, around a mean value of  $\sim 23 \mu s$. Such a value is fully consistent with previous measurements \citep{vaughan98}. 
\subsection{Frequency dependent time lags}
Across its full frequency span, the QPO cannot be detected on timescale as short of 32 seconds in the two adjacent bands. For this reason, we have adopted 128 second integration time for the lag measurements {(i.e. averaging $M=32$ intervals of 4 seconds)}. In Table \ref{dbt1} we list the observations considered in this paper, together with the mean lag measured. We have split the data in ObsIDs first, but also within an ObsIDs in the contiguous segments of data provided by the {\it science event} files. In deriving the mean soft lag, we have considered only segments of 128 seconds, in which the QPO is detected above $\sim 4.5 \sigma$ (excess power) in both the soft and hard energy bands (this corresponds to a cut of $R \ge 1.5$, where $R$ is the ratio between the Lorentzian normalization and its $1 \sigma$ error, see \cite{boutelier10} for a discussion). For a given integration time, the QPO significance depends on the source count rate and frequency; the latter parameter defining the QPO RMS amplitude and width \citep[][]{barret05}. In the dataset considered, the significance of the QPO goes down at both ends of the frequency range, more dramatically at the lower end where the source count rate was the lowest. Nevertheless, adopting the cuts above, enables to detect the lags between $\sim 565$ and $\sim 890$ Hz. In Figure \ref{dbf2}, we show the soft lags measured on 128 seconds as a function of the QPO frequency. The lags have also been binned into ten adjacent frequency bins. A clear trend is now obvious in the data, in particular above 680 Hz where the soft lags smoothly decrease with frequency. 

\subsection{The energy spectrum of the lags}
We now wish to examine how the lags vary with energy. In each energy bin, {using the same procedure as described above}, the  lag is computed between the light curve in that bin and the light curve in the reference energy band (3-30 keV), where the signal to noise ratio of the QPO detection is the highest. The light curve in the energy bin considered is subtracted from the reference light curve, to ensure that Poisson noise remains uncorrelated \citep{uttley11}. For each segment of 128 seconds, { hence for each QPO frequency $\nu_i$}, for each energy bin, we thus have a measurement of the lag, { as the average of $M=32$ values}. A mean lag-energy spectrum can be computed by averaging all the measurements { weighted by their errors}, within a continuous {\it science event} file duration, and even within one ObsID, provided that the channel energy boundaries do not change. Two examples of such lag spectra are shown in Figure in \ref{dbf3} for the first two ObsIDs which provide the better statistics (10072-05-01-00 and 30062-02-01-000). Note that the data were unfortunately not recorded in the same spectral mode in the two ObsIDs: 64, 32 channels. Although of limited significance, for the first ObsID where the source count rate is the highest, it is interesting to note the lag spectrum is fairly steep below 4 keV, and seems to flatten around 5-8 keV, where the weak and broad Iron line is significantly present in the energy spectrum (see below), before steepening again. There is also a dip in the lag spectrum just around 6.4 keV. { This could be due to the presence of a constant and narrow iron line component (e.g. the core of the broad line), produced at larger distances from the neutron star or over a wider range of radii than its red wing}. { The same features are however not so clear in the other segment of data shown in Fig. \ref{dbf3} (right panel), although the data were recorded with a poorer spectral resolution, which could easily hide the structure seen at higher spectral resolution.}
\subsection{Time averaged continuum emission spectrum}
In order to place the results described above into context, it is worth looking at the time averaged energy spectrum to see how it changes with respect to the QPO frequency. One of the most extensive spectral analysis of the RXTE data, presented in \cite{gierlinski02} sampled the full set of spectral states of the source (it did not include any timing information though). When grouping data based on spectral colors (as to improve the statistics), they showed that across the Z-shaped color-color diagram, all X-ray spectra could be represented as the sum of a disk blackbody component, plus a comptonized component and a weak reflection component from an ionized disk. Both the reflection and its self consistently evaluated iron line emission were statistically detected in all the spectra. In this paper, we have extracted {\it Standard 2} spectra for all the time intervals of Table \ref{dbt1}. We have combined the data from all PCUs units and used the latest calibration data available in {\it heasoft 6.12}, adding a systematic uncertainly of 0.25\% to the data. We have frozen the column density to $1.5 \times 10^{22}$ cm$^{-2}$ \citep{penninx89}. Although there is more than one possible decomposition of the X-ray spectrum of \4u, \cite[e.g.][]{lin07,takahashi11}, we have assumed the widely used spectral model of \cite{gierlinski02} \citep[see also, e.g.][for a discussion of models]{barret00,barret01,disalvo02,tarana08,cackett10}. However, with the reduced statistical quality of our data compared to \cite{gierlinski02}, we were unable to obtain stable fits when adding the self-evaluted reflection of the Comptonized emission (this is largely a consequence of having two curvy components and a broad iron line profile within the relatively limited bandpass of the PCA). We have also used the {\it Comptt} comptonization model \citep{titarchuk94} instead of the {\it nthcomp} model \citep{zycki99}. We have checked however that both models provide consistent  fits. The addition of a {\it diskline} component improved the fit significantly in all segments of data, as previously found \citep{gilfanov03,gierlinski02}. Unfortunately, the data do not allow to constrain the inner disk radius of the {\it diskline} model, which we have frozen to $10 R_g$ (for a disk inclination of 45 degrees). Different values (e.g.  $6 R_g$) would fit the data as well. The normalization ($N_{DBB}$) of the {\it diskbb} component was converted into an inner disk radius ($R_{in}$), following  \cite{gierlinski02}:
\begin{equation}
R_{in} = 0.61 N_{DBB}^{1/2} \left(\frac{2.7}{\eta}\right) \left(\frac{D}{3.6~kpc}\right) \left( \frac{f_{col}}{1.8}\right)^2 \left(\frac{0.5}{cos~i}\right)^{1/2}~km
\end{equation}
where $D$ is the source distance, $i$ is the disk inclination angle, $f_{col}$ is the ratio of the color to effective temperature \citep{shimura95} and $\eta$ is the correction factor for the inner torque-free boundary condition ($\eta = 2.7$ for $R_{in} = 6R_g$)Ê\citep{gierlinski99}. 

The best fit spectral parameters are listed in Table \ref{dbt2} and plotted in Figure \ref{dbf4} against the lower kHz QPO frequency. As can be seen, all spectral parameters show a smooth behavior with frequency, suggesting that our spectral decomposition is robust. From table 2, the significance of the line can be inferred from looking at the $\chi^2$ with and without the line. In our observations, the $\Delta \chi^2$ varies between $\sim 15$ up to $\sim 60$ for 2 additional degrees of freedom (and a total of degrees of freedom around 40), making the {\it diskline}
component, highly significant. 
\section{Summary of the results}
\begin{figure}
\plotone{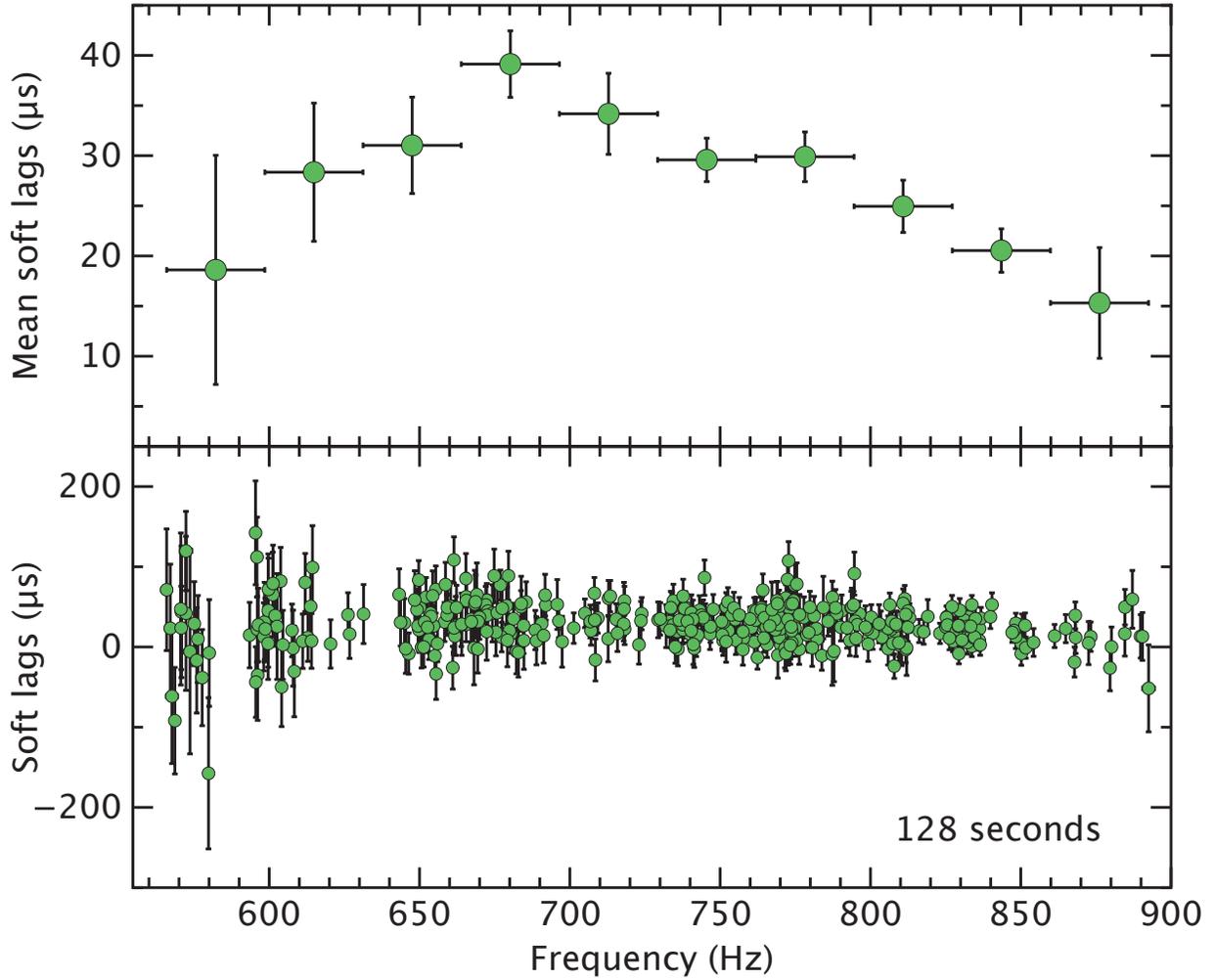}
\caption{Top) Soft lags between the 3--8 keV and 8--30 keV time series, binned in 10 adjacent QPO frequency intervals. The lags show a clear dependency with frequency, with the most significant trend being the smooth decrease at frequencies above $\sim 650$ Hz. Bottom) Same as above, but the individual soft lags measured on 128 second timescales are shown. \label{dbf2}}
\end{figure}

\begin{figure}
\epsscale{1.10}
\plottwo{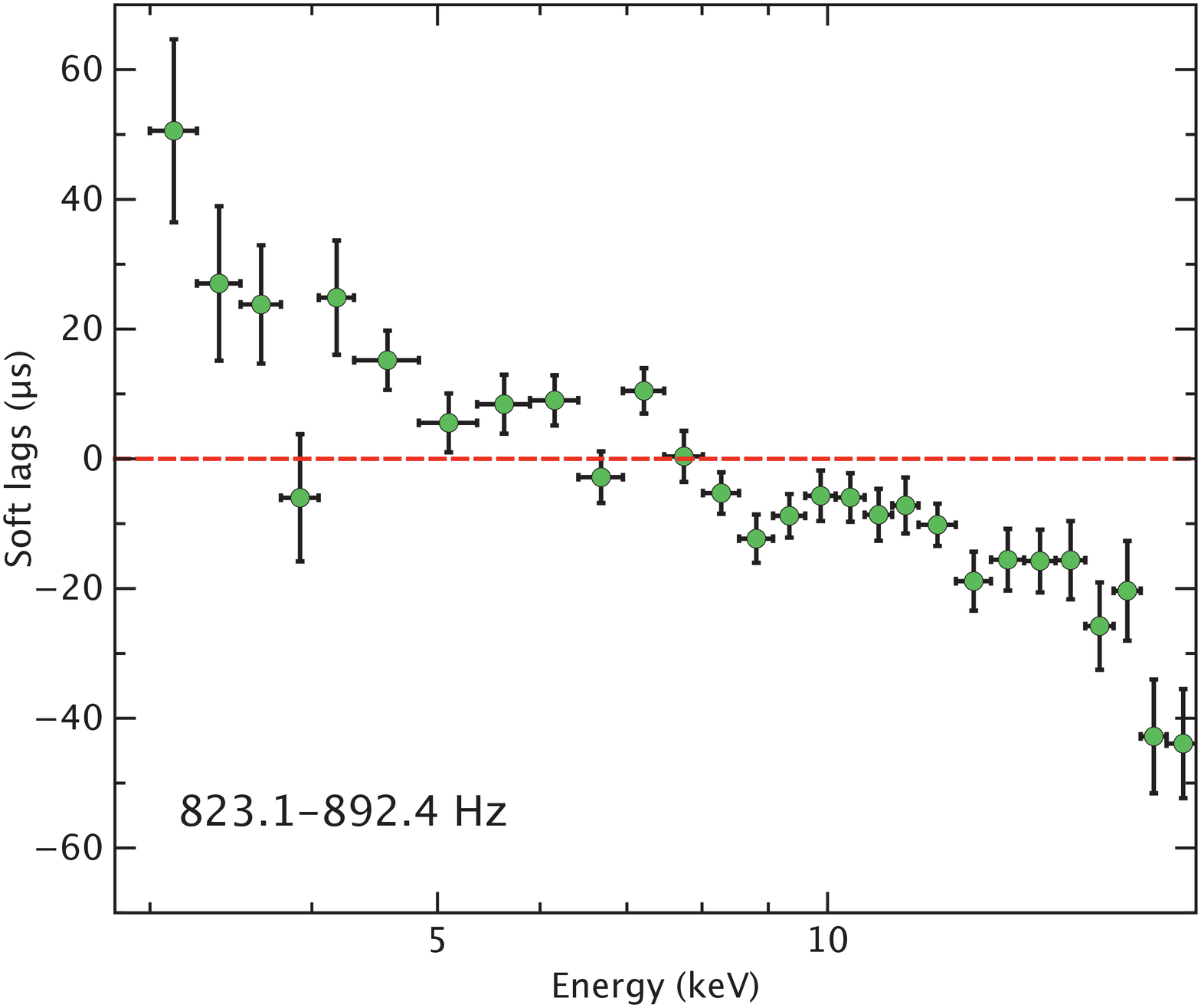}{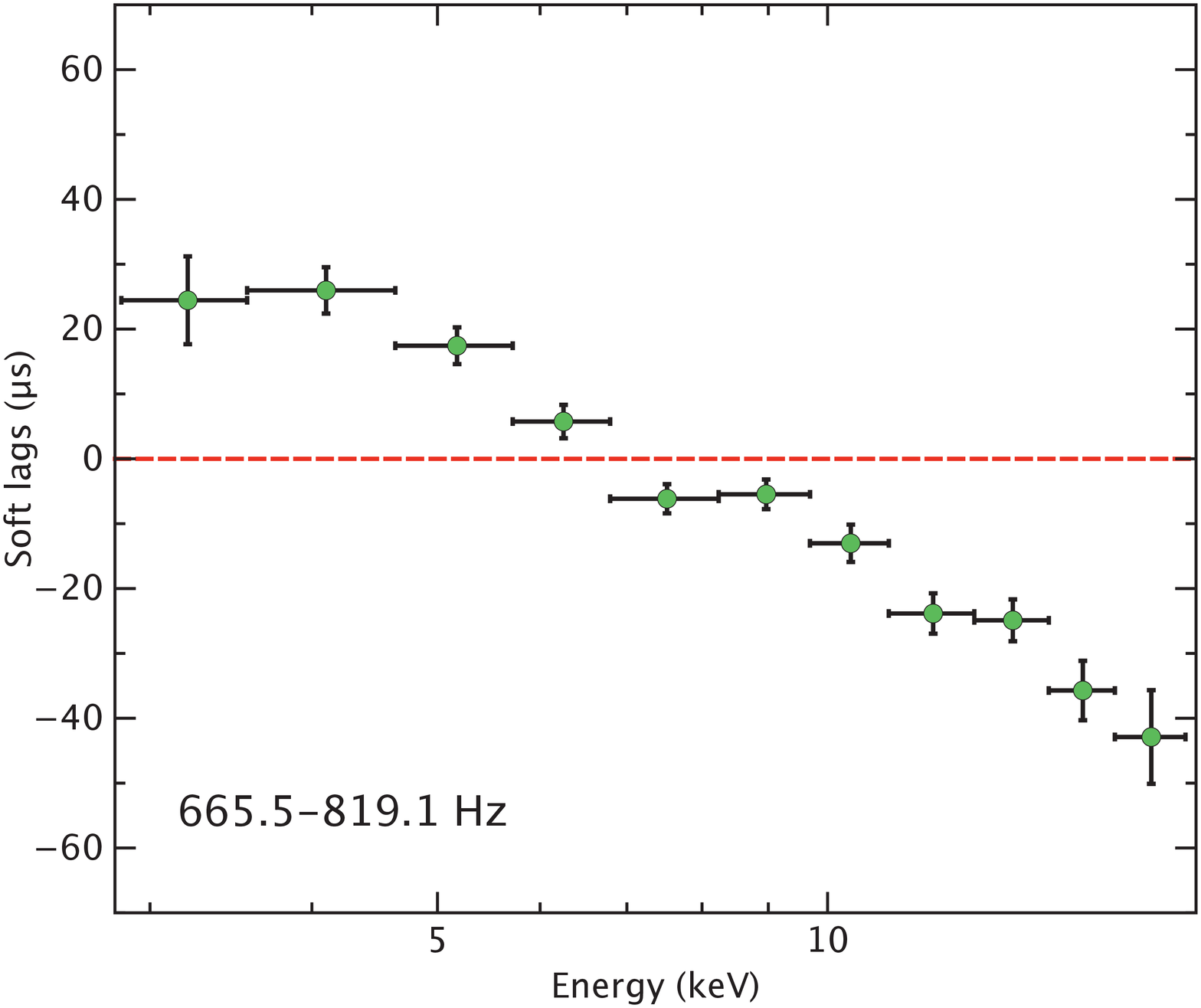}
\caption{Mean soft lag energy spectra measured for the ObsIDs 10072-05-01-00 (left) and 30062-02-01-000 (right). The lag-energy spectrum on the left is the best that could be obtained, as a combination of a narrow spectral binning and high count rate statistics. Note in particular the broad bump between 5 and 7 keV, suggestive of iron line reverberation. The dip around 6.4 keV may be associated with a constant and narrow iron line component (e.g. the core of the broad line), produced at larger distance from the neutron star or over a wider range of radii than the red wing of the line. It is also worth noting that the mean lag around 15 keV is -20 $\mu s$ while the lag around 5-7 keV is about 10 $\mu s$. The total lag of 30 $\mu s$ corresponds to a distance of $\sim 4 R_g$ for a 1.4 M$_\odot$ neutron star. For a neutron star of 10 km or $5 R_g$ radius, this would put the inner disk radius at $\sim 9 R_g$; a value that would fit the data with the {\it diskline} model and that is typically inferred from fitting the broad iron line of neutron stars \citep{cackett10}. However, the trend seen on the left hand side plot is not so obvious on the plot on the right hand side, but note that the data were recorded with a much lower spectral resolution. \label{dbf3}}
\end{figure}

\begin{figure}
\epsscale{1.10}
\plottwo{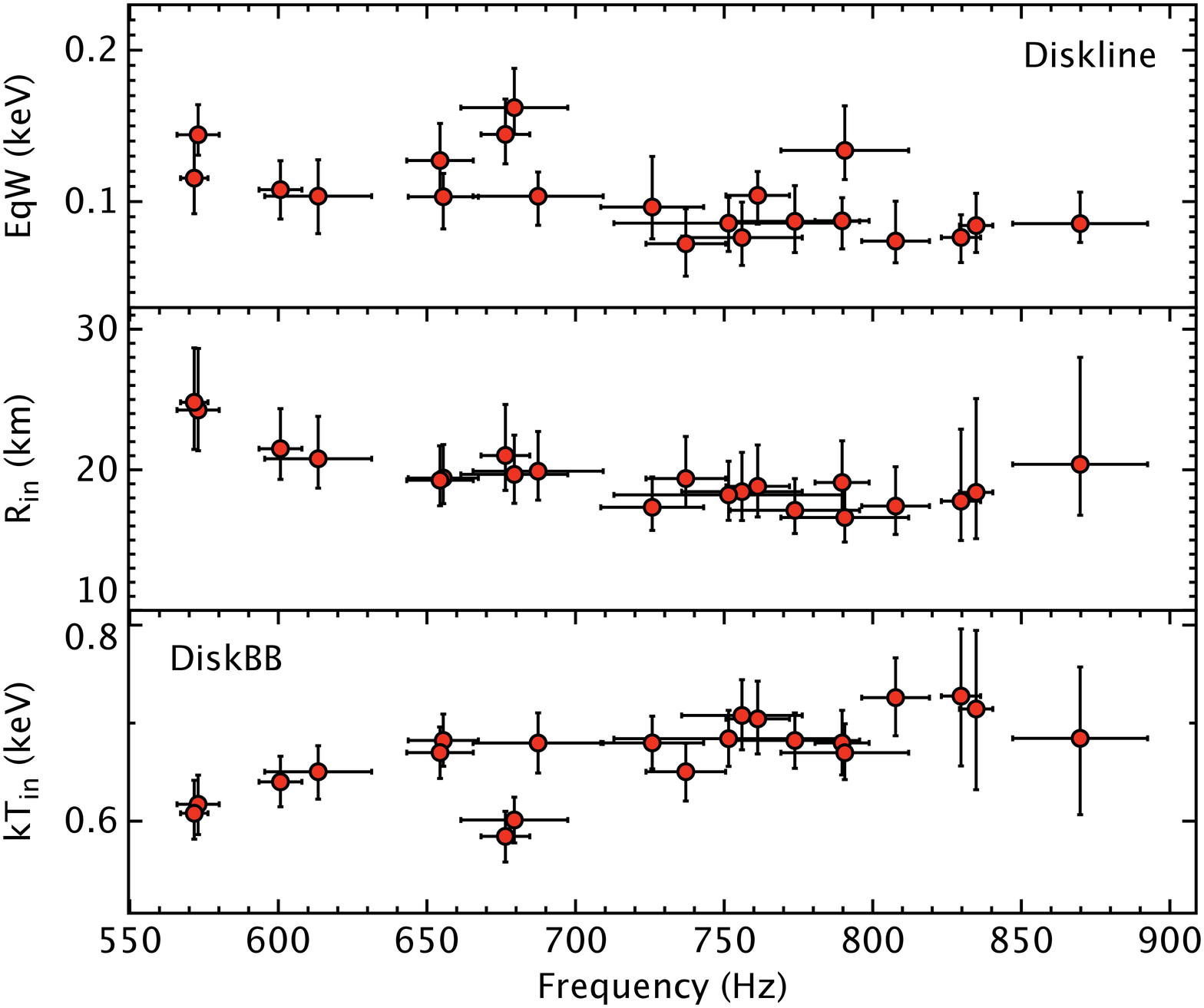}{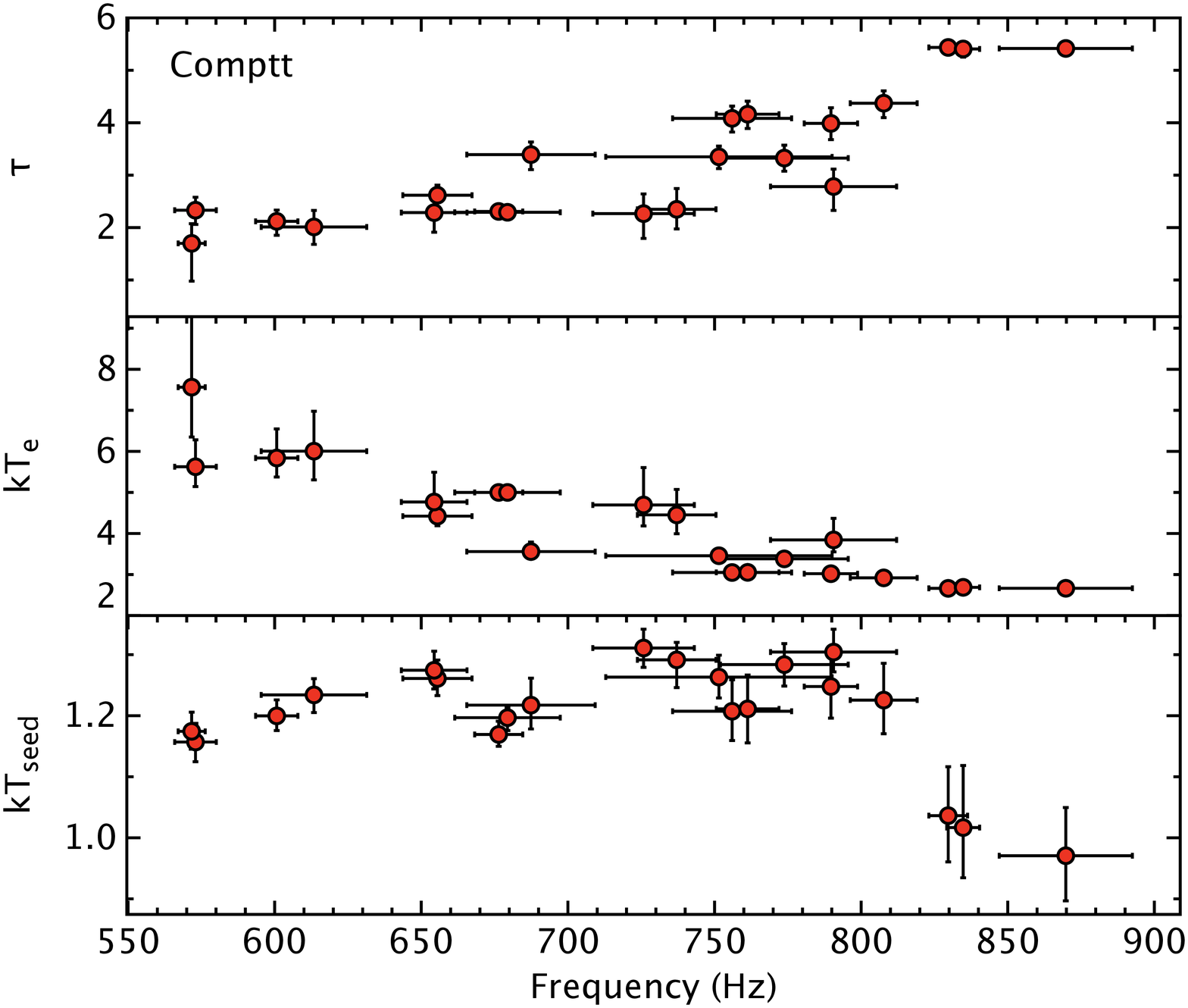}
\caption{Variation of the best fit spectral parameters as a function of the lower kHz QPO frequency. Left) From top to bottom, the iron line equivalent width (in keV), the inner disk radius (in km), computed from equation 1 (with canonical parameters), and the inner disk temperature (in keV). Right) From top to bottom, the optical depth, the electron temperature (in keV) and the seed photon temperature (in keV). Errors on the fitted parameters are given at the $1\sigma$ level. The error bar on the frequency is determined by the spread of QPO frequency over the continuous {\it science event} file. As can be seen, all best fit spectral parameters, but the seed photon temperature (at the highest frequencies), show a smooth behavior with frequency.  \label{dbf4}}
\end{figure}

\begin{figure}
\epsscale{0.8}
\plotone{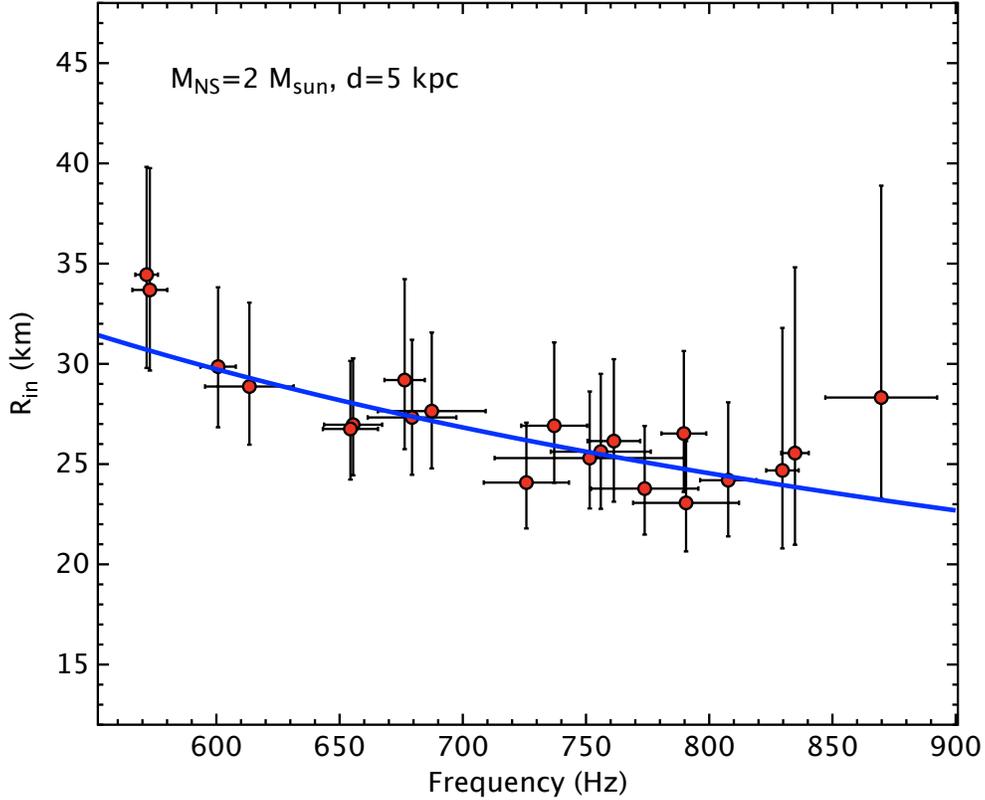}
\caption{Same as Figure \ref{dbf4} but the inner disk radius inferred from spectral fitting is computed from equation 1 for a distance of 5 kpc. The solid line represents the orbital radius derived from the orbital frequency (assumed to be provided by the lower kHz QPO) around a neutron star mass of 2 M$_\odot$. The Spearman correlation coefficient is -0.69, corresponding to a null hypothesis probability of $\sim 5\times10^{-4}$. The source distance is  $5.8^{+2.0}_{-1.9}$ kpc according to \cite{guver10}. \label{dbf5}}
\end{figure}

As advocated for instance by \cite{uttley11}, frequency and energy dependent time lags give additional information, compared to time average energy and power density spectra, providing new and independent constraints on models for the emission mechanisms and the location at which they take place in the system. Our main results can be summarized as follows:
\begin{itemize}

\item The soft lags between the 3-8 and 8-30 keV QPO photons vary with frequency, dropping at both end of the frequency range investigated (from 565 to 890 Hz). In particular, the lags decrease smoothly from about 40 $\mu s$ to 15 $\mu s$, while the frequency varies from 680 Hz to 890 Hz. Our results are consistent with the values reported by \cite{avellar13} below 800 Hz. The same curvy shape in the lag-frequency plot seems to be present in the data of 4U1636-536 \citep[see Figure 3 top panel][]{avellar13}, suggesting that this may be a common feature of lags from lower kHz QPOs. It should be stressed however that this is not the intrinsic lags that is measured, due to the fact primary direct and delayed emissions are likely to contribute to the emission in both energy bands \citep[the so-called lag dilution, e.g.][]{zoghbi11}. With this caveat in mind, it is still worth trying to relate the change in the lags with frequency with proxies of the disk location, such as the kHz QPO frequencies or the inner disk radii inferred from the spectral fitting of the continuum emission.

\item When the lower kHz QPO frequency varies from $\sim 600$ to $\sim 900$ Hz, the upper kHz QPO frequency varies from $\sim 900$ to $\sim 1070$ Hz. For a neutron star of 1.4-2.0 M$_\odot$, assuming that either frequency is an orbital frequency at the inner edge of the disk, this implies a change of the inner disk radius of $\sim 6.2-7.0$ km { (or $\sim 3.5 R_g$ for a 1.4 M$_\odot$ canonical neutron star)} or $2.2-2.4$ km, or equivalently a light travel time difference of 23-21 or 7-8 $\rm \mu$s. These values are not very different from the span of the soft lags measured. Interestingly, considering only the decreasing part of the lag-frequency variation of Fig. \ref{dbf1} (between 680 and 900 Hz), the relative change of the lags by $\sim 25 \mu$s would be very comparable to the light travel time difference, associated with the relative change of the orbital radius inferred from the lower kHz QPO frequency change (4.7 km or $\sim 16~\mu$s). 

\item The spectral analysis of the continuum emission is consistent with a picture in which the disk gets closer to the neutron star, while the QPO frequency increases. The relative change of the inner disk radius inferred from fitting the disk component is again broadly consistent with the span of the lags measured. It is also consistent with the relative radius change assuming that the lower kHz QPO is an orbital frequency. The distance to the source is not very well known, ranging from 3.6 kpc \citep{nakamura89}, as used by \citep{gierlinski02} in equation 1, to a more recent value of  $5.8^{+2.0}_{-1.9}$ kpc \citep{guver10} from type I burst analysis \citep[see also][]{suleimanov11}). Assuming for instance a distance of 4.4 kpc or 5.0 kpc in equation 1, for a neutron star of 1.4 M$_\odot$ or 2.0 M$_\odot$ respectively, would provide a perfect match between the inferred inner disk radius variations from spectral fitting and the orbital radius changes assuming the lower kHz QPO is providing the orbital frequency (see Fig. \ref{dbf5}). On the other hand, this would not work if one assumed that the upper kHz QPO is providing an orbital frequency, because the change in radii associated with the change of frequency would be too limited ($\sim 2$ km as opposed to $\sim 6-7$ km). We note that the debate on which one of the two kHz QPO frequencies is an orbital frequency is not yet settled \citep{osherovich99,sanna12}, although most models predict that it is the upper QPO that provides an orbital frequency \citep[e.g.][]{miller98,klis00}. { As discussed by \cite{lamb01}, it is also possible that the observed upper kHz QPO frequency is significantly lower than an orbital frequency, which may then span a wider frequency range and reach higher frequencies, so that the associated change of radius would be actually larger than observed. In absolute units (km), to make it consistent with the inner disk radius changes inferred from spectral fitting (Fig. \ref{dbf4}), one would need to tune some of the parameters in equation 1, so that $R_{in}$ drops by some 20\% or so. This does not seem unfeasible given the uncertainties on the four parameters of equation 1 (e.g. the source distance).} As a side note, interpreting the drop of the quality factor of the lower kHz QPO, assumed to be orbital, as a signature of the innermost stable circular orbit would then be more natural \citep{barret06}. However, it would be necessary to explain how the system can generate frequencies higher than the ISCO frequency (i.e. the upper kHz QPO). This would also imply large masses for the neutron star in kHz QPO sources. In the case of 4U1608-522, if the ISCO frequency is at 900 Hz (the highest frequency detected here), then the gravitational neutron star mass would be around 2.6 M$_\odot$, assuming the dimensionless angular momentum of the star ($j\equiv cJ/(GM^2)$) to be 0.1 \citep{miller98}. Such a value would be significantly larger than the mass of the most massive neutron star known to date \citep{demorest10}. 

\item Evidence for reflection comes from the presence of a broad ionized Iron line, significantly detected in all spectra, although weak, as pointed out by \cite{gierlinski02}. This is a necessary ingredient for the reverberation scenario. The presence of several components (from the disk, the comptonization cloud and the reflection) in the PCA bandpass, together with its limited spectral resolution do not enable us to extract the parameters of the reflection component (e.g. ionization state of the reflector). There is a trend for the equivalent width of the line to decrease with frequency that needs to be understood.

\item Understanding the lag energy spectrum would require detailed simulations of the transfer function of the irradiated disk, but looking at Figure \ref{dbf2}, it is quite striking that they may already contain some very valuable information. This is particularly true in the highest count rate observations (recorded with the highest spectral resolution), which show both a soft excess and a broad bump around the broad iron line, present in the averaged energy X-ray spectrum. It is thus tempting to interpret the soft excess as thermal reprocessing of the hard X-ray photons onto the disk, and the bump around the broad iron line as the signature of delayed reflection. { Further support for this interpretation comes from looking at the energy spectrum of the lag \ref{dbf2}. The mean lag around 15 keV is -20 $\mu s$ while the lag around 5-7 keV is about 10 $\mu s$. The total lag of 30 $\mu s$ corresponds to a distance of $\sim 4 R_g$ for a 1.4 M$_\odot$ neutron star. For a neutron star of 10 km or $5 R_g$ radius, this would put the inner disk radius at $\sim 9 R_g$; a value that would fit the data with the {\it diskline} model and that is typically inferred from fitting the broad iron line of neutron stars \citep{cackett10}}. More detailed modeling is still required to test the reverberation scenario. 
\end{itemize}

\section{Conclusions}
To conclude, although alternative explanations for the soft lags have been proposed, for instance, as being due to an intrinsic spectral softening of the emission along the QPO cycle \citep{kaaret99}, or due to temperature oscillations in the corona driving with some delays the temperature oscillations of the soft photon source \citep{lee01,avellar13}, the recent detection of soft lags in a variety of compact objects, from stellar mass black holes to AGNs, and the evidence accumulated above suggest that the lags measured involves reverberation of the hard pulsating source located in the boundary layer and interacting with the accretion disk. It is worth noting that for a similar geometry, down scattering of hard X-rays (from the accretion column) in a cool medium (the accretion disk) was proposed to explain the $\sim 100\mu s$ soft lags observed in the pulsed emission of millisecond pulsars \citep[e.g.][]{falanga07}. In our data, the distance between the irradiating hard source and the reflector varies in a way that can be tracked by the frequency of kHz QPOs.

Our results illustrate clearly the power of lag measurements to probe the innermost regions of the accretion flows onto compact objects.
Even if the surfaces and faster variability timescales of neutron stars make them more complex to analyze than black holes, kHz QPO
sources offer stable and coherent clocks that allow us to measure time lags comparable to the light crossing time of a few R$_g$ (the light crossing time at $1 R_g$ is about $7 \mu s$ for a canonical neutron star), just
as for AGN. Interpreting the detailed shape of the lag energy spectra, and determining intrinsic lag values will require modeling the transfer function of the response of the accretion disk to the irradiating source, which is beyond the scope of this paper. Putting a yardstick on a neutron star accretion disk will then be within reach.
The joint timing and spectral analysis presented here should also
be extended to other sources for which kHz QPOs were detected cleanly
with the RXTE PCA  \citep[such as 4U1636-536,][]{avellar13}.  The
discovery of patterns similar to what we have found for \4u\ (e.g. broad bump around the iron line in the energy spectrum of the lags) would add support to the reverberation scenario.  It would also prepare the ground for breakthrough observations to be performed by the next generation of timing missions,
such as LOFT \citep{feroci11}, whose combination of large
effective area and improved spectral resolution will uncover exquisite
details about the soft lags for both the lower and upper kHz QPOs, providing errors on the lags twenty time smaller and enabling truly iron line reverberation studies. This holds great potential to constrain the neutron star mass and to understand the underlying physics of the QPO emission and its propagation in the strong field region around the neutron star.

\section*{Acknowledgments}

It is my pleasure to thank Phil Uttley for very stimulating discussions, that triggered the analysis presented here. Great thanks also to Cole Miller, Jean-Pierre Lasota, Ed Cackett, Jon Miller, Chris Done and Jean-Francois Olive for useful discussions along the preparation of this paper. {The author wishes to thank an anonymous referee whose comments helped to clarify some of the statements made in the paper.}

\clearpage

\clearpage

\clearpage

\begin{deluxetable}{lccccclc}
\tabletypesize{\scriptsize}
\rotate
\tablecaption{Log of lags detections for \4u on timescale of 128 seconds. The table lists the observation date, the ObsID, the number of the {\it science event} file within the ObsID, the start date of the observation data set, the frequency span, the mean source count rate, the mean time lag and associated error over the {\it science event} file considered, the $\chi^2$ and the  number of degrees of freedom for a fit by a constant, and finally the exact energy bands used for the soft band (the first two energies), the hard band (the second and third energies).\label{dbt1}}
\tablewidth{0pt}
\tablehead{
\colhead{ObsID} & \colhead{n} & \colhead{Start date} & \colhead{Frequency space} & \colhead{Count rate} &
\colhead{Mean lag} & $\chi^2$ (d.o.f) & \colhead{Energy bands} 
}
\startdata
10072-05-01-00&1&1996/03/03--19:17:21&847.2--892.5&2943.3&14.3$\pm3.3$&21.5~(24)&3.0--8.0--30.5 \\
10072-05-01-00&2&1996/03/03--20:53:19&823.1--836.3&2827.9&22.6$\pm2.4$&24.2~(27)&3.0--8.0--30.5 \\
10072-05-01-00&3&1996/03/03--22:29:18&829.3--840.4&2861.9&25.1$\pm4.3$&19.3~(13)&3.0--8.0--30.5 \\
30062-02-01-000&1&1998/03/24--15:36:34&750.7--772.0&1954.0&27.7$\pm3.3$&16.4~(20)&2.9--8.2--32.4 \\
30062-02-01-000&2&1998/03/24--17:06:23&796.4--819.1&2042.9&23.9$\pm3.8$&36.1~(26)&2.9--8.2--32.4 \\
30062-02-01-000&3&1998/03/24--18:42:22&735.7--776.3&1943.7&25.2$\pm3.5$&27.8~(26)&2.9--8.2--32.4 \\
30062-02-01-000&4&1998/03/24--20:18:20&665.5--709.3&1818.4&38.0$\pm3.5$&19.1~(25)&2.9--8.2--32.4 \\
30062-02-01-00&1&1998/03/24--22:01:31&643.7--667.2&1751.6&26.1$\pm4.9$&19.5~(23)&2.9--8.2--32.4 \\
30062-02-01-00&2&1998/03/24--23:53:30&643.2--665.6&1415.0&40.3$\pm8.6$&19.4~(15)&2.9--8.2--32.4 \\
30062-02-01-01&1&1998/03/25--13:54:37&593.5--607.9&1547.6&26.7$\pm9.4$&15.3~(18)&2.9--8.2--32.4 \\
30062-02-01-01&2&1998/03/25--15:30:25&565.9--580.0&1531.1&12.2$\pm29.5$&12.1~(8)&2.9--8.2--32.4 \\
30062-02-01-01&3&1998/03/25--17:46:60&567.1--576.3&1467.5&11.1$\pm11.1$&1.3~(6)&2.9--8.2--32.4 \\
30062-02-01-01&4&1998/03/25--18:42:22&595.4--631.4&1512.4&29.3$\pm8.9$&10.5~(13)&2.9--8.2--32.4 \\
30062-02-01-02&3&1998/03/26--15:30:25&712.9--790.2&1569.2&32.1$\pm3.8$&27.0~(26)&2.9--8.2--32.4 \\
30062-02-01-02&4&1998/03/26--17:06:23&780.6--798.9&1638.3&30.6$\pm6.2$&15.7~(12)&2.9--8.2--32.4 \\
30062-01-01-00&1&1998/03/27--12:27:36&708.5--743.1&1332.3&26.5$\pm4.9$&7.7~(12)&2.9--8.6--30.9 \\
30062-01-01-00&3&1998/03/27--13:53:23&723.6--750.5&1312.7&34.3$\pm4.3$&15.3~(19)&2.9--8.6--30.9 \\
30062-01-01-00&5&1998/03/27--15:29:21&752.0--795.6&1393.1&35.2$\pm5.1$&33.8~(27)&2.9--8.6--30.9 \\
30062-01-01-01&1&1998/03/28--14:02:52&769.1--812.0&1152.2&28.2$\pm7.3$&22.5~(16)&2.9--8.6--30.9 \\
30062-01-01-02&2&1998/03/29--09:05:16&668.1--684.6&916.6&36.9$\pm6.6$&5.3~(11)&2.9--8.6--30.9 \\
30062-01-01-02&4&1998/03/29--10:41:15&661.5--697.4&911.2&39.1$\pm8.6$&17.9~(15)&2.9--8.6--30.9 \\
\enddata
\end{deluxetable}

\begin{deluxetable}{lccccccccclc}
\tabletypesize{\scriptsize}
\rotate
\tablecaption{Best fit spectral parameters listed by ObsID and {\it event files}, numbered $n$, within an ObsID. The parameters are: the normalization of the disk blackbody temperature kT$_{in}$, the normalization N$_{DBB}$ of the disk blackbody component, the seed photon temperature kT$_{seed}$, the electron temperature kT$_e$, the optical depth of the comptonization cloud $\tau$, the line energy of the {\it diskline} component, its equivalent width, the 2-20 keV X-ray luminosity ($L_x$), the reduced $\chi^2$, with and without the {\it diskline} component. All errors are quoted at the $1\sigma$ level. \label{dbt2}}
\tablewidth{0pt}
\tablehead{
\colhead{ObsID} & \colhead{n} & \colhead{kT$_{in}$} & \colhead{N$_{DBB}$} & \colhead{kT$_{seed}$}  &\colhead{kT$_e$} &  \colhead{$\tau$}  & \colhead{E} & \colhead{EqW} & \colhead{$L_x$} &\colhead{$\chi^2$} &\colhead{$\chi^2$} 
}
\startdata
10072-05-01-00 & 1 & $0.68^{+0.07}_{-0.08}$ & $1580.00^{+1179.41}_{-561.27}$ & $0.97^{+0.08}_{-0.08}$ & $2.66^{+0.03}_{-0.02}$ & $5.43^{+0.13}_{-0.11}$ & $7.27^{+0.14}_{-0.13}$ & $85.46^{+22.65}_{-16.81}$ & $9.01 \pm 0.01$ & 29.54 (44) & 58.74 (46) \\ 
10072-05-01-00 & 2 & $0.73^{+0.07}_{-0.07}$ & $1200.28^{+666.12}_{-378.47}$ & $1.03^{+0.08}_{-0.08}$ & $2.66^{+0.03}_{-0.03}$ & $5.45^{+0.13}_{-0.13}$ & $7.32^{+0.18}_{-0.17}$ & $76.34^{+19.62}_{-12.24}$ & $8.66 \pm 0.01$ & 30.36 (44) & 53.11 (46) \\ 
10072-05-01-00 & 3 & $0.71^{+0.08}_{-0.08}$ & $1285.93^{+933.07}_{-460.49}$ & $1.02^{+0.10}_{-0.10}$ & $2.69^{+0.04}_{-0.03}$ & $5.42^{+0.15}_{-0.15}$ & $7.38^{+0.14}_{-0.15}$ & $84.20^{+21.75}_{-19.92}$ & $8.80 \pm 0.01$ & 23.46 (44) & 46.62 (46) \\ 
30062-02-01-000 & 1 & $0.70^{+0.04}_{-0.04}$ & $1347.54^{+420.20}_{-312.27}$ & $1.21^{+0.06}_{-0.06}$ & $3.04^{+0.13}_{-0.10}$ & $4.20^{+0.26}_{-0.29}$ & $7.31^{+0.17}_{-0.16}$ & $104.11^{+16.14}_{-23.87}$ & $5.97 \pm 0.01$ & 31.71 (37) & 61.23 (39) \\ 
30062-02-01-000 & 2 & $0.73^{+0.04}_{-0.04}$ & $1153.56^{+370.36}_{-267.42}$ & $1.23^{+0.06}_{-0.06}$ & $2.92^{+0.11}_{-0.08}$ & $4.38^{+0.24}_{-0.28}$ & $6.99^{+0.17}_{-0.16}$ & $73.91^{+22.86}_{-12.70}$ & $6.25 \pm 0.01$ & 22.43 (37) & 38.50 (39) \\ 
30062-02-01-000 & 3 & $0.71^{+0.04}_{-0.03}$ & $1294.09^{+391.00}_{-289.14}$ & $1.23^{+0.06}_{-0.05}$ & $3.06^{+0.13}_{-0.10}$ & $4.06^{+0.24}_{-0.28}$ & $7.09^{+0.18}_{-0.18}$ & $76.20^{+20.74}_{-16.21}$ & $5.93 \pm 0.01$ & 27.60 (37) & 45.18 (39) \\ 
30062-02-01-000 & 4 & $0.68^{+0.03}_{-0.03}$ & $1505.44^{+427.17}_{-311.83}$ & $1.22^{+0.04}_{-0.04}$ & $3.59^{+0.21}_{-0.16}$ & $3.35^{+0.23}_{-0.26}$ & $7.25^{+0.14}_{-0.13}$ & $103.55^{+19.75}_{-16.67}$ & $5.55 \pm 0.01$ & 35.12 (37) & 67.42 (39) \\ 
30062-02-01-00 & 1 & $0.68^{+0.03}_{-0.03}$ & $1432.42^{+351.69}_{-268.61}$ & $1.26^{+0.03}_{-0.03}$ & $4.41^{+0.29}_{-0.23}$ & $2.63^{+0.19}_{-0.21}$ & $7.05^{+0.14}_{-0.12}$ & $103.21^{+24.50}_{-14.98}$ & $5.35 \pm 0.01$ & 42.08 (44) & 73.16 (46) \\ 
30062-02-01-00 & 2 & $0.67^{+0.03}_{-0.03}$ & $1409.94^{+357.80}_{-266.24}$ & $1.28^{+0.03}_{-0.03}$ & $4.75^{+0.68}_{-0.44}$ & $2.30^{+0.31}_{-0.37}$ & $6.76^{+0.14}_{-0.11}$ & $127.15^{+25.05}_{-20.53}$ & $4.24 \pm 0.01$ & 57.42 (44) & 93.62 (46) \\ 
30062-02-01-01 & 1 & $0.64^{+0.03}_{-0.03}$ & $1757.95^{+464.67}_{-357.54}$ & $1.20^{+0.03}_{-0.03}$ & $5.80^{+0.65}_{-0.44}$ & $2.14^{+0.20}_{-0.25}$ & $7.06^{+0.07}_{-0.13}$ & $107.87^{+25.70}_{-13.31}$ & $4.74 \pm 0.01$ & 36.26 (44) & 70.81 (46) \\ 
30062-02-01-01 & 2 & $0.62^{+0.03}_{-0.03}$ & $2235.57^{+806.66}_{-533.78}$ & $1.16^{+0.03}_{-0.04}$ & $5.60^{+0.65}_{-0.53}$ & $2.35^{+0.28}_{-0.27}$ & $7.29^{+0.15}_{-0.13}$ & $144.15^{+25.85}_{-20.59}$ & $4.71 \pm 0.01$ & 43.39 (44) & 91.01 (46) \\ 
30062-02-01-01 & 3 & $0.61^{+0.03}_{-0.03}$ & $2337.68^{+730.09}_{-631.76}$ & $1.17^{+0.04}_{-0.03}$ & $7.25^{+3.98}_{-1.05}$ & $1.79^{+0.34}_{-0.74}$ & $7.43^{+0.19}_{-0.19}$ & $115.49^{+35.85}_{-14.21}$ & $4.53 \pm 0.01$ & 31.74 (44) & 56.95 (46) \\ 
30062-02-01-01 & 4 & $0.65^{+0.03}_{-0.03}$ & $1641.90^{+476.01}_{-330.28}$ & $1.23^{+0.03}_{-0.03}$ & $5.96^{+0.96}_{-0.69}$ & $2.03^{+0.32}_{-0.33}$ & $7.09^{+0.16}_{-0.14}$ & $103.60^{+27.48}_{-17.48}$ & $4.62 \pm 0.01$ & 50.44 (44) & 78.31 (46) \\ 
30062-02-01-02 & 3 & $0.68^{+0.03}_{-0.03}$ & $1264.76^{+328.37}_{-254.71}$ & $1.26^{+0.04}_{-0.03}$ & $3.45^{+0.16}_{-0.13}$ & $3.37^{+0.21}_{-0.22}$ & $7.19^{+0.18}_{-0.17}$ & $85.64^{+23.20}_{-21.31}$ & $4.74 \pm 0.01$ & 32.55 (44) & 54.79 (46) \\ 
30062-02-01-02 & 4 & $0.68^{+0.03}_{-0.03}$ & $1388.04^{+428.23}_{-306.72}$ & $1.24^{+0.05}_{-0.05}$ & $3.00^{+0.15}_{-0.11}$ & $4.03^{+0.29}_{-0.32}$ & $7.25^{+0.24}_{-0.22}$ & $87.23^{+24.31}_{-22.82}$ & $4.94 \pm 0.01$ & 59.54 (44) & 77.47 (46) \\ 
30062-01-01-00 & 1 & $0.68^{+0.03}_{-0.03}$ & $1142.19^{+283.54}_{-217.18}$ & $1.31^{+0.03}_{-0.03}$ & $4.65^{+0.92}_{-0.50}$ & $2.30^{+0.37}_{-0.48}$ & $7.22^{+0.20}_{-0.17}$ & $96.42^{+31.71}_{-18.28}$ & $4.00 \pm 0.01$ & 52.70 (44) & 72.65 (46) \\ 
30062-01-01-00 & 3 & $0.65^{+0.03}_{-0.03}$ & $1426.96^{+440.14}_{-290.18}$ & $1.28^{+0.04}_{-0.04}$ & $4.28^{+0.59}_{-0.39}$ & $2.48^{+0.35}_{-0.39}$ & $7.34^{+0.44}_{-0.36}$ & $72.18^{+19.32}_{-17.30}$ & $3.93 \pm 0.01$ & 42.92 (44) & 57.14 (46) \\ 
30062-01-01-00 & 5 & $0.68^{+0.03}_{-0.03}$ & $1114.06^{+292.50}_{-215.30}$ & $1.28^{+0.04}_{-0.04}$ & $3.37^{+0.18}_{-0.14}$ & $3.35^{+0.24}_{-0.26}$ & $7.14^{+0.19}_{-0.16}$ & $86.96^{+20.24}_{-18.39}$ & $4.19 \pm 0.01$ & 39.73 (44) & 61.13 (46) \\ 
30062-01-01-01 & 1 & $0.67^{+0.03}_{-0.03}$ & $1048.10^{+280.18}_{-220.29}$ & $1.31^{+0.04}_{-0.04}$ & $3.86^{+0.52}_{-0.32}$ & $2.77^{+0.38}_{-0.45}$ & $7.29^{+0.11}_{-0.11}$ & $133.80^{+24.05}_{-23.26}$ & $3.42 \pm 0.01$ & 38.99 (44) & 81.70 (46) \\ 
30062-01-01-02 & 2 & $0.58^{+0.03}_{-0.03}$ & $1678.76^{+579.16}_{-396.59}$ & $1.17^{+0.02}_{-0.02}$ & 5.00 & $2.31^{+0.02}_{-0.02}$ & $7.57^{+0.18}_{-0.16}$ & $144.39^{+23.61}_{-21.08}$ & $2.68 \pm 0.00$ & 51.30 (45) & 93.63 (47) \\ 
30062-01-01-02 & 4 & $0.60^{+0.02}_{-0.02}$ & $1471.02^{+417.69}_{-308.22}$ & $1.20^{+0.02}_{-0.02}$ & 5.00 & $2.29^{+0.02}_{-0.02}$ & $7.46^{+0.12}_{-0.11}$ & $162.04^{+23.74}_{-16.79}$ & $2.67 \pm 0.00$ & 45.07 (45) & 109.00 (47) \\ 
\enddata
\end{deluxetable}

\end{document}